\documentclass[preprint,12pt]{elsarticle}

\usepackage[T1]{fontenc}
\usepackage[utf8]{inputenc} 
\usepackage{lmodern}
\usepackage{amsmath,amssymb}
\usepackage{graphicx}
\usepackage[none]{hyphenat}
\usepackage{xcolor}
\usepackage{hyperref}

\usepackage{subfig} 

\journal{Chaos, Solitons \& Fractals}

\begin{document}

\begin{frontmatter}

\title{Complex transitions between spiking, bursting and silent regimes in a new memristive Rulkov neuronal model}

\author[aff1]{Miguel~Moreno}
\author[aff1]{Alexandre~R.~Nieto}
\author[aff1,aff2]{Miguel~A.~F.~Sanju\'an}

\address[aff1]{Nonlinear Dynamics, Chaos, and Complex Systems Group, Departamento de F\'isica, Universidad Rey Juan Carlos, M\'ostoles, 28933, Madrid, Spain}
\address[aff2]{Royal Academy of Sciences of Spain, Valverde 22, 28004 Madrid, Spain}

\begin{abstract}
The Rulkov model, which simulates the behavior of biological neurons, is modified by replacing one of its control parameters with a memristive, sigmoid-type function of finite memory. This modification causes the parameter to vary according to the system’s history throughout the simulation. Previous works usually modify the Rulkov model by introducing additional parameters altering its behavior. Here, by contrast, we retain the original equations and allow the control parameters to vary in time, thereby preserving the model’s fundamental properties. In this sense, the proposed model is locally equivalent in time to the original one. However, unlike the original model, which reproduces a single neuronal regime per simulation, the new memristive version exhibits both uniform and chaotic transitions among multiple neuronal activity regimes. Its dynamics are examined with respect to the rate at which the memristive function changes and the number of internal states it stores. Three distinct scenarios emerge around a bifurcation point. Before the bifurcation, the system undergoes uniform transitions toward a stable bursting regime. After the bifurcation, it shows uniform transitions toward a final spiking or silent regime. At the bifurcation point, highly complex transitions arise. As examples, we present trajectories in which the neuron chaotically switches between regimes without ever settling, and trajectories for which it requires around $140000$ map iterations to reach a stationary regime.
\end{abstract}


\end{frontmatter}

\section{Introduction}
Since the proposal of the Hodgkin-Huxley model \cite{HodgkinHuxley} in 1952, most mathematical descriptions of neuronal activity relied on systems of coupled ordinary differential equations involving the membrane potential as one of the state variables. This paradigm shifted with the pioneering work of Chialvo \cite{Chialvo} in 1995, who introduced a neuronal model based on a two-dimensional discrete-time dynamical system, or map. The Chialvo model was followed by several map-based models (see Ref.~\cite{MapModels} for a complete review), among which the Rulkov models~\cite{Rulkov2001, Rulkov2002, Rulkov2004} stand out as some of the most effective in reproducing the behavior of biological neurons. Despite their reduced dimensionality, map-based models are capable of generating complex dynamics that would otherwise require higher-dimensional continuos-time models. Even one-dimensional maps can exhibit multistability and chaos \cite{SinusoidalMap}. Furthermore, map-based models significantly reduce the computational cost of simulating large neuronal networks.\\

Recently, diverse attempts have been made to enhance the Rulkov model by introducing a memristive function, that is, a non-constant parameter whose value depends on the accumulated sum of one of the state variables, inspired by Chua’s theoretical discovery of the memristor \cite{Chua}. These memristive extensions of the Rulkov model exhibit extreme multistability \cite{Ding, Multistability} and have been shown to be effective both as pseudorandom number generators \cite{Bao1, Bao2} and for implementing hybrid digital watermarking embedding schemes \cite{Ding}. However, they diverge from the author's original goal of reproducing physiologically observed neuronal features.\\

In addition to the conventional approach, new strategies have been developed to incorporate memristive schemes into maps, such as the step-wise coupling method \cite{StepWise}, which relies on the combined action of multiple memristive functions. Most memristive maps assume infinite memory, meaning that the memristor state is determined by an indefinite sum of past states. In contrast, in this work we introduce a memristive function with finite memory. This choice gives rise to an additional important nonlinear effect, namely the replacement of old states by new ones, and enables the dynamics of the model to be controlled by the memory length. As a consequence, memory acts as a control parameter of the system, and the resulting dynamics depend on the number of stored states.\\

This paper is structured as follows. In Sec.~\ref{sec: Original Rulkov model}, we review the main properties of the original Rulkov model. In Sec.~\ref{sec: Memristive Rulkov model}, we introduce the new memristive Rulkov model and examine its dynamics, identifying a bifurcation that alters both the type and the number of stable neuronal regimes, shifting from a single bursting regime to two possible regimes, spiking or silent. Finally, in Sec.~\ref{sec: Conclusions}, we summarize the main findings of this work.

\section{Overview of the original Rulkov model}
\label{sec: Original Rulkov model}
\begin{figure}[h!]
	\centering
	\includegraphics[width=\linewidth]{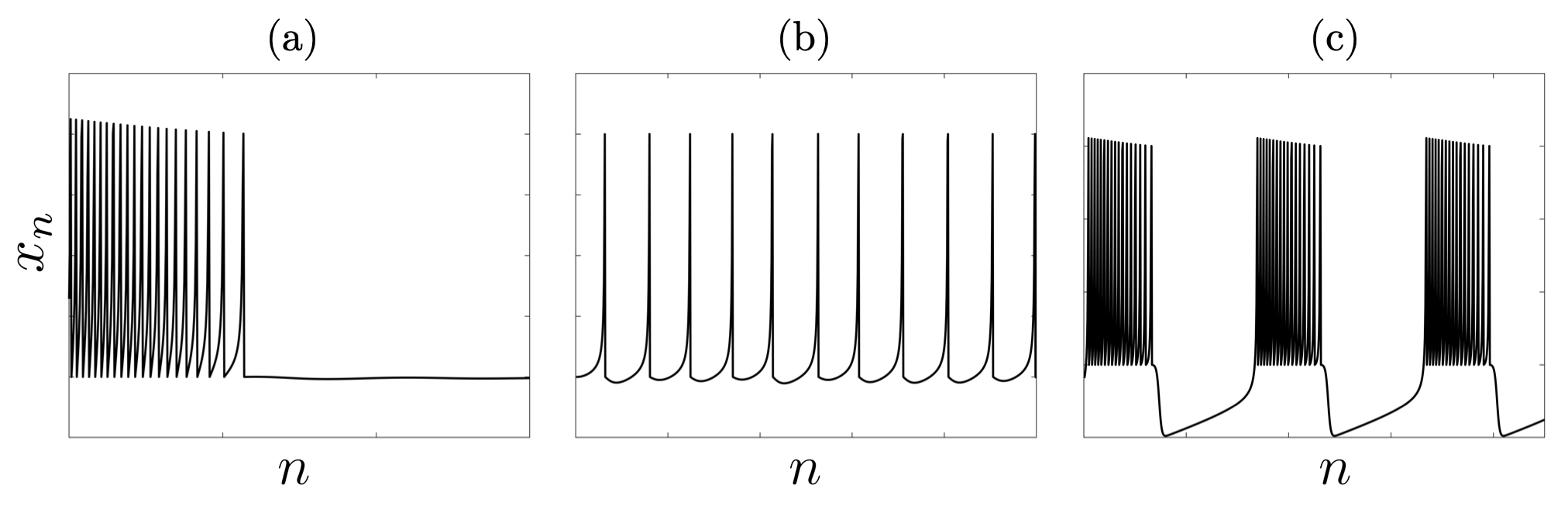}
	\caption{Examples of the three neuronal regimes: (a) silent, (b) spiking, and (c) bursting, where $x$ represents the neuron's membrane voltage and $n$ indicates the time.}
	\label{fig: NeuronalRegimes}
\end{figure}

Rulkov proposed three simple models that effectively reproduce the main regimes observed in biological neurons \cite{Rulkov2001, Rulkov2002, Rulkov2004}: silent, spiking, and bursting (see Figs.~\ref{fig: NeuronalRegimes}(a)–(c)). These models are based on a two-dimensional map of the form

\begin{equation}
\left \{
\begin{array}{l}
x_{n+1}=f(x_n,y_n),\\
y_{n+1}=y_n-\mu(x_n+1-\sigma),\\
\end{array}
\right .
\label{eq: Rulkov}
\end{equation}
\vspace{0.1cm}

\noindent and differ only in the specific expression of the function $f(x,y)$. In Eq.~(\ref{eq: Rulkov}), $x$ represents the neuron's membrane voltage, $y$ is an auxiliary variable without a direct physical interpretation, and $\sigma$ and $\mu$ are control parameters. The parameter $\mu$ is set to a small positive value, making $x$ the fast variable and $y$ the slow variable. In this study, we fix $\mu = 0.001$ and focus on the model proposed by Rulkov in 2002 \cite{Rulkov2002}, in which the fast map is given by

\begin{align}
    f(x,y)=\begin{cases}
        \dfrac{\alpha}{1-x}+y &; \hspace{0.25cm} x\leq 0,\\
        \alpha+y &; \hspace{0.25cm} 0<x<\alpha+y,\\
        -1 &; \hspace{0.25cm} x\geq\alpha+y.
        \end{cases}
\label{eq: Rulkov2002}
\end{align}
\vspace{0.1cm}

Rulkov used Eq.~(\ref{eq: Rulkov2002}) to reproduce the spikes that occur within bursts, a feature absent from his earlier model \cite{Rulkov2001}. With this modification, he achieved a frequent bursting regime consistent with experimental observations (see Fig. 2 in Ref.~\cite{BurstingFeatures}), in which both the amplitude and frequency of oscillations decrease toward the end of each burst, and the bursts occur at higher membrane voltages than the silent phases. Nevertheless, other types of bursting are also possible (see, for instance, Fig.~2 in  Ref.~\cite{TwoDucks}). According to the bursting classification proposed by Izhikevich and Hoppensteadt in Ref.~\cite{BurstingClassification}  based on co-dimension-$1$ bifurcations, the Rulkov model produces ``fold/homoclinic''-type bursting.\\

\begin{figure}[h!]
	\centering
	\includegraphics[width=0.8\linewidth]{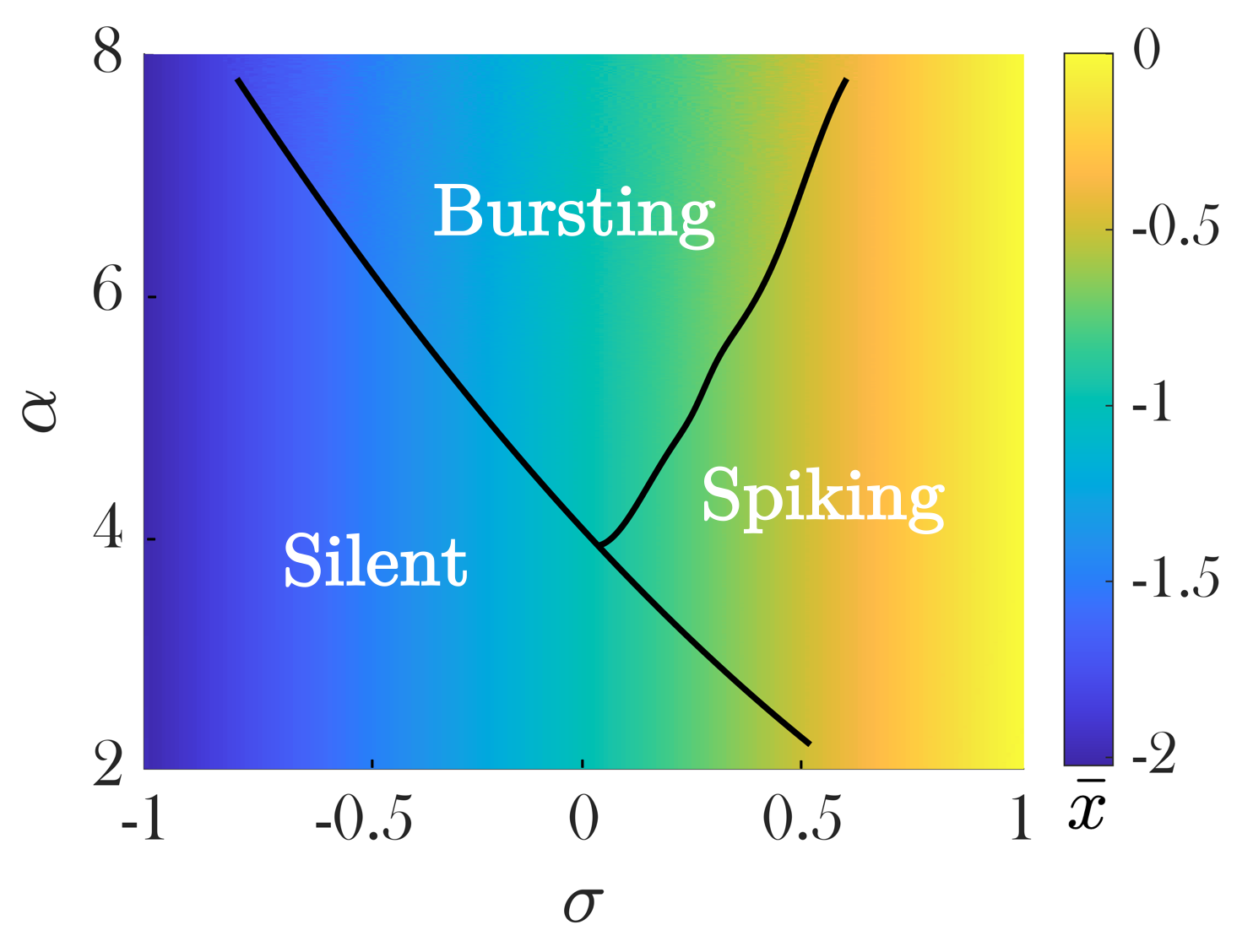}
	\caption{The $(\sigma,\alpha)$ plane is divided into regions according to the neuronal regime obtained for each pair of control-parameter values in the original Rulkov model \cite{Rulkov2002}. The background color represents the mean value of the fast variable, $\bar{x}$.}
	\label{fig: ParameterPlane}
\end{figure}

For each pair of control-parameter values $(\sigma, \alpha)$, the original 2002 model generates one of the three neuronal regimes, as illustrated in Fig.~\ref{fig: ParameterPlane}. This figure shows the $(\sigma,\alpha)$ parameter plane divided into three regions according to the resulting neuronal activity. The background is colored by the mean value of the fast variable $\bar{x}$, a quantity that will be useful in the following section.\\

\begin{figure}[h!]
	\centering
	\includegraphics[width=\linewidth]{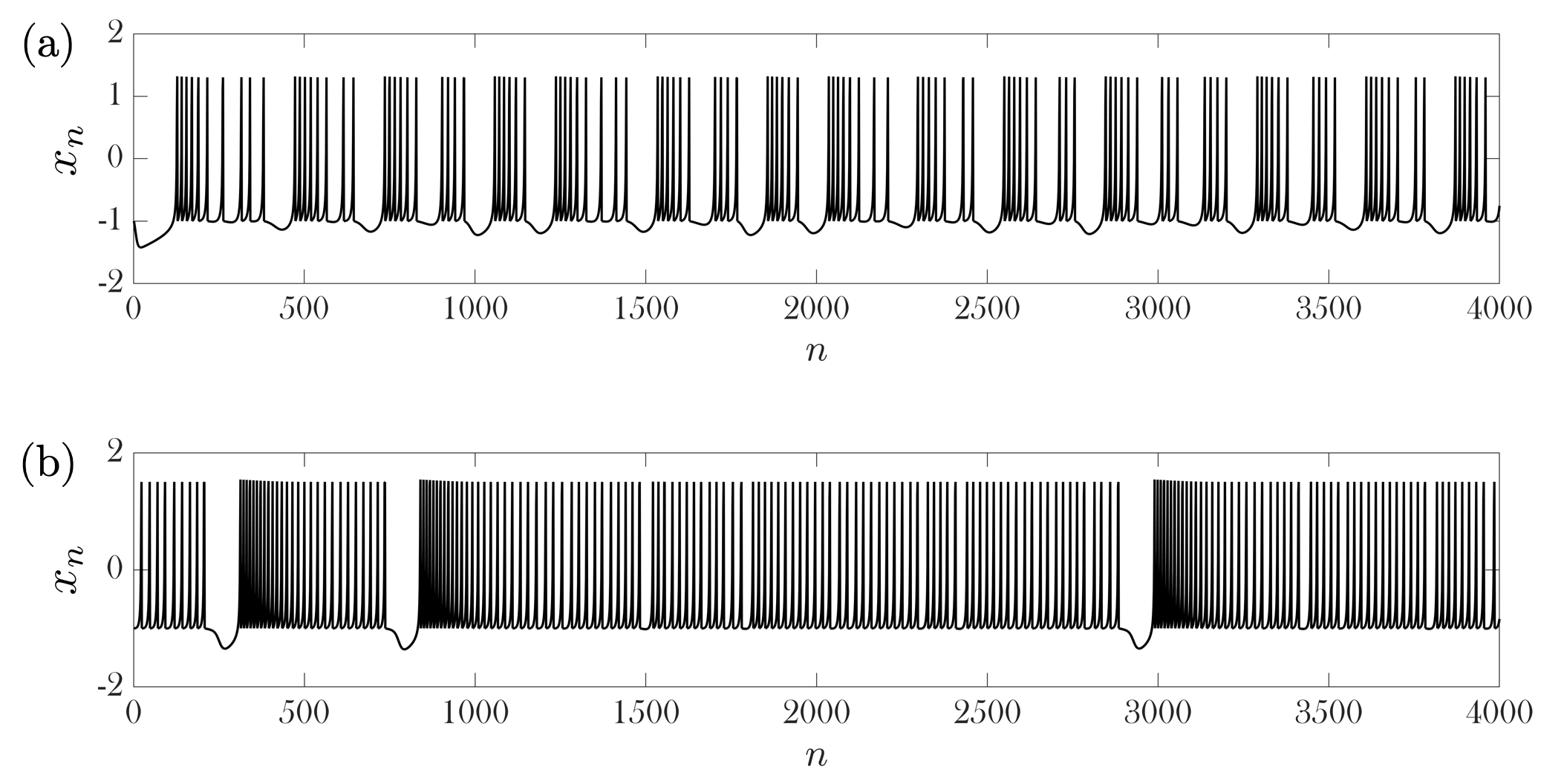}
	\caption{Transitions between spiking and bursting regimes using the original Rulkov model \cite{Rulkov2002} for (a) $\alpha=4.6$, $\sigma=0.16$, $\mu=0.001$ and $(x_0,y_0)=(-1,-3.33)$ as initial condition, and for (b) $\alpha=5$, $\sigma=0.28$, $\mu=0.001$ and $(x_0,y_0)=(-1,-3.5)$ as initial condition. Transitions are rather unclear and may also be interpreted as irregular bursting.}
	\label{fig: OriginalTransitions}
\end{figure}

The objective of this research is to modify the Rulkov model, governed by Eqs.~(\ref{eq: Rulkov}) and (\ref{eq: Rulkov2002}), so that transitions between neuronal regimes can occur within a single simulation based on the system’s own history. Although it has been shown \cite{ChaosRulkov} that the original model can also exhibit chaotic transitions near the regions boundaries in the $(\sigma,\alpha)$ plane, such transitions are typically very subtle and make it difficult to clearly distinguish the corresponding neuronal regimes. Figs.~\ref{fig: OriginalTransitions}(a) and \ref{fig: OriginalTransitions}(b) illustrate transitions between spiking and bursting in the original model for different parameter values. These transitions are rather unclear and may also be interpreted as irregular bursting. In contrast, the transitions obtained with the memristive model proposed here are significantly more complex, and the distinctions between neuronal regimes become much more pronounced.

\section{Memristive Rulkov model}
\label{sec: Memristive Rulkov model}
Instead of adding a new term to incorporate a memristive mechanism into the model, as done in previous works, we replace the control parameters $\alpha$ and $\sigma$ with memristive functions, thereby preserving the structure of the original Rulkov model at each time step. This ensures that the new model continues to reproduce the biological behavior it was designed to represent. With this strategy, the control parameters become time-dependent, and the system is able to cross the boundaries between regions in Fig.~\ref{fig: ParameterPlane}, producing transitions between different neuronal regimes. Following an approach similar to that used in \cite{MemristiveHenon} for the Hénon map, the new memristive functions are given by

\begin{subequations}
\begin{align}
	\alpha(z)&=\alpha_{-}+(\alpha_{+}-\alpha_{-})\varphi(z), \label{eq: MemristiveAlpha}\\
	\sigma(z)&=\sigma_{-}+(\sigma_{+}-\sigma_{-})\varphi(z), \label{eq: MemristiveSigma}
\end{align}
\end{subequations}
\vspace{0.1cm}

\noindent where $z$ is the memristive state variable and $\varphi(z)$ is the sigmoid function 

\begin{equation}
	\varphi(z)=\frac{1}{1+\exp{(-z/\tau)}},
	\label{eq: SigmoidFunction}
\end{equation}
\vspace{0.1cm}

\noindent which guarantees that $\alpha$ and $\sigma$ remain confined to the intervals $(\alpha_{-},\alpha_{+})$ and $(\sigma_{-},\sigma_{+})$, respectively, where the subscripts $\pm$ denote the limiting values as $z\rightarrow\pm\infty$. In Eq.~(\ref{eq: SigmoidFunction}), the constant $\tau$ controls the rate at which the memristive function changes. We consider a discrete memristor with a finite memory of $m$ states, similar to the one proposed in \cite{FiniteMemoryMemristor}, where the memristive variable is given by

\begin{equation}
	z_n=\begin{cases}
        		z_0 +\sum\limits_{i=0}^{n-1}(x_i+h) &; \hspace{0.25cm} 0\leq n\leq m,\\
        		 \sum\limits_{i=n-m}^{n-1}(x_i+h) &; \hspace{0.25cm} n > m.
        		\end{cases}
	\label{eq: FiniteMemory}
\end{equation}
\vspace{0.1cm}

Note that in Eq.~(\ref{eq: FiniteMemory}), an additional term $h > 0$ is included to address the issue that arises when $x_n < 0$ constantly. As shown in Fig.~\ref{fig: ParameterPlane}, the mean value of the fast variable is negative for almost any choice of $\sigma$ and $\alpha$, making this correction necessary. Hereafter, we refer to the Rulkov model that uses Eq.~(\ref{eq: MemristiveSigma}) (or Eq.~(\ref{eq: MemristiveAlpha})) in place of the original constant parameter as the $\sigma$-memristive (respectively, $\alpha$-memristive) model.\\

Using $\alpha(z)$ as a memristive function turns out to be ineffective for two main reasons. First, the dynamics of the $\alpha$-memristive model are completely predictable because the mean value $\bar{x}$ barely changes when moving vertically in the parameter plane, as Fig.~\ref{fig: ParameterPlane} reveals. Second, in the original model, the amplitude of the spikes is given by $x=\alpha+y$ and the evolution of $y$ is properly designed to recreate the main experimental results found during the bursting. Therefore, allowing $\alpha$ to vary disrupts this mechanism, leading to biologically unrealistic bursting regimes in which the amplitude no longer decreases toward the end of each burst. In contrast, the capabilities of the model are significantly enhanced when using $\sigma(z)$ as a memristive function according to Eq.~(\ref{eq: MemristiveSigma}). This corresponds to moving horizontally across the control-parameter plane (see Fig.~\ref{fig: ParameterPlane}). Unlike the $\alpha$-memristive case, such horizontal movement produces substantive changes in $\bar{x}$, leading to nontrivial dynamical behaviors.\\

In the following, we discuss the dynamics of the $\sigma$-memristive Rulkov model in three possible scenarios. First, we explore the dynamics prior to a bifurcation, where the model reproduces a stable bursting regime. Next, we examine the dynamics after the bifurcation, where the system settles into either a spiking regime or a silent regime, depending on the initial conditions. Finally, we analyze the dynamics of the model at the bifurcation, revealing complex transitions among neuronal regimes.

\subsection{Dynamics of the $\sigma$-memristive Rulkov model prior to the bifurcation}
\begin{figure}[h!]
	\centering
	\includegraphics[width=\linewidth]{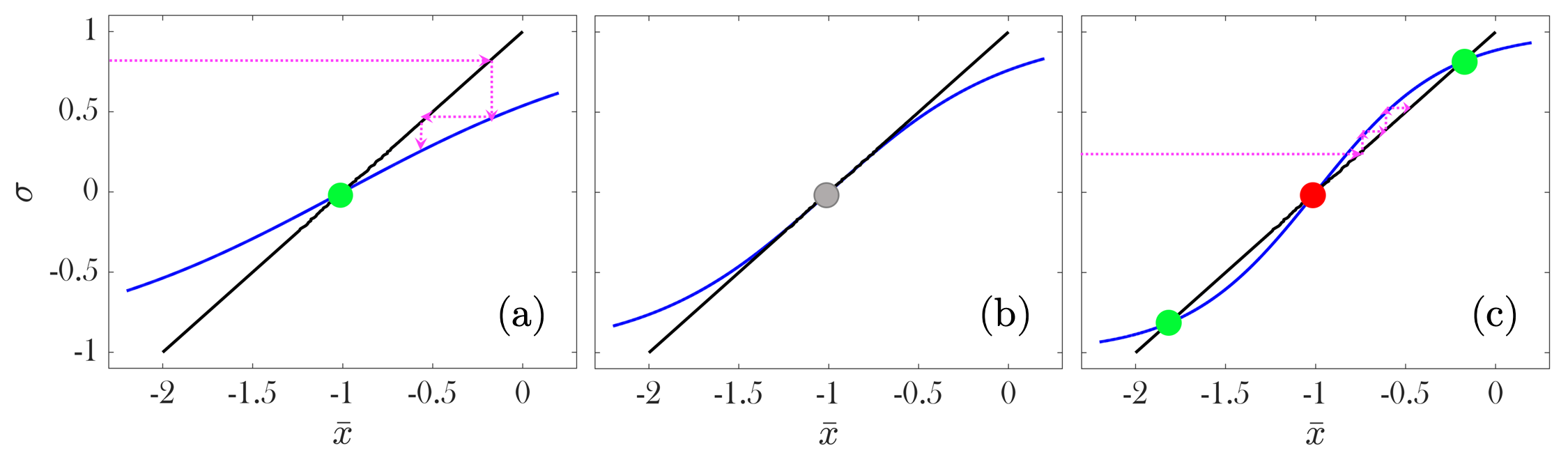}
	\caption{The black line $\sigma_0$ represents the relation between the control parameter $\sigma$ and the mean value $\bar{x}$ in the original Rulkov model \cite{Rulkov2002}, given by Eq.~(\ref{eq: MeanValueVsSigma}). The blue curves represent the memristive function $\sigma(\bar{x})$, given by Eq.~(\ref{eq: MeanMemristiveSigma}) for $\tau=50$ and different values of the number of states stored by the memristor: (a) $m=60$ (prior to the bifurcation), (b) $m=100$ (at the bifurcation), and (c) $m=140$ (after the bifurcation). The intersection between the curves is indicated by a green dot when it corresponds to a stable neuronal regime, and by a red dot otherwise.}
	\label{fig: Bifurcation}
\end{figure}

For the $\sigma$-memristive model, we set $\sigma_{-}=-1$, $\sigma_{+}=1$ and $h=1$ in Eqs.~(\ref{eq: MemristiveSigma}) and (\ref{eq: FiniteMemory}), based on the horizontal limits and the mean value of the fast variable depicted in Fig.~\ref{fig: ParameterPlane}. We also fix the control parameter $\alpha=5$, so that the three neuronal regimes are equally possible, as Fig.~\ref{fig: ParameterPlane} suggests. For these particular values, Eq.~(\ref{eq: MemristiveSigma}) becomes

\begin{equation}
	\sigma(\bar{x})=-1+\frac{2}{1+\exp[{-(m/\tau)(\bar{x}+1)]}},
	\label{eq: MeanMemristiveSigma}
\end{equation}
\vspace{0.1cm}

\noindent where the memristive variable $z$, given by Eq.~(\ref{eq: FiniteMemory}), has been replaced by the mean value of the fast variable $x$. We expect the new memristive model to enter a stationary neuronal regime when the memristive function $\sigma(\bar{x})$ converges to a value that leads to a dynamics where the mean value $\bar{x}$ equals that of the original model. These stationary regimes correspond to the intersection points between the black line and the blue curve in Fig.~\ref{fig: Bifurcation}. The black line represents the mean value $\bar{x}$ for each value of the control parameter $\sigma$ in the original Rulkov model, which is given almost exactly by

\begin{equation}
	\sigma_0:\hspace{0.25cm} \sigma=\bar{x}+1,
	\label{eq: MeanValueVsSigma}
\end{equation} 
\vspace{0.1cm}

\noindent while the blue curve represents the new memristive function, given by Eq.~(\ref{eq: MeanMemristiveSigma}). The bifurcation occurs when $m\approx2\tau$. In Fig.~\ref{fig: Bifurcation}(a), both $\sigma(\bar{x})$ and $\sigma_0$ are plotted prior to the bifurcation (i.e., $m<2\tau$), particularly for $m=60$ and $\tau=50$, where they only intersect once at the point $(\bar{x},\sigma)=(-1,0)$, which corresponds to a stable regime of the $\sigma$-memristive model. At the bifurcation, the curve $\sigma(\bar{x})$ is tangent to the line $\sigma_0$ at $\bar{x}=-1$, as depicted in Fig.~\ref{fig: Bifurcation}(b) for $m=100$ and $\tau=50$. After the bifurcation (i.e., $m>2\tau$), the regime located at $(\bar{x},\sigma)=(-1,0)$ becomes unstable and two new intersections between the curves appear, both corresponding to stable regimes, as shown in Fig.~\ref{fig: Bifurcation}(c) for $m=140$ and $\tau=50$.

\begin{figure}[h!]
	\centering
	\includegraphics[width=\linewidth]{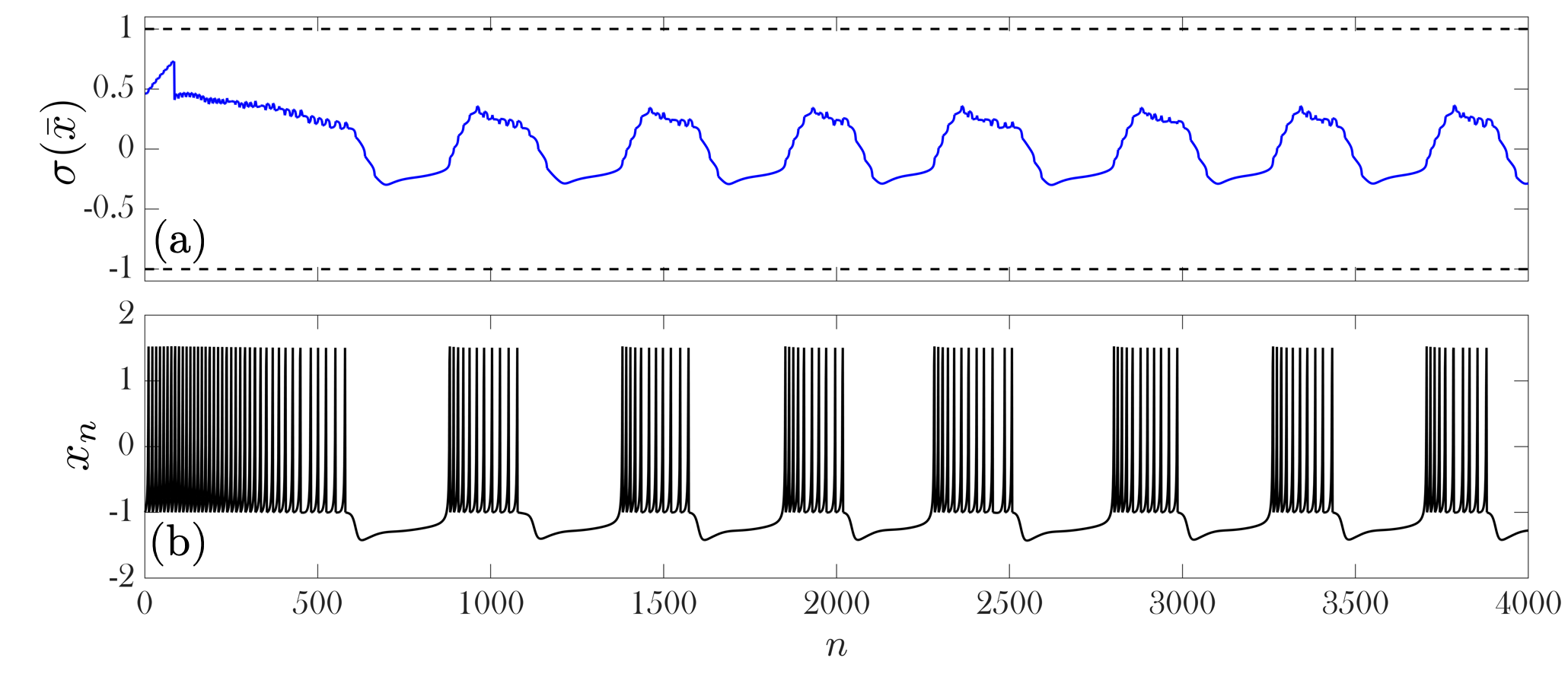}
	\caption{Evolution, prior to the bifurcation, of the memristive function $\sigma(\bar{x})$ (a), given by Eq.~(\ref{eq: MeanMemristiveSigma}), and the state variable $x$ (b) for $\alpha=5$, $\mu=0.001$, $\tau=50$, $m=85$ and $(x_0,y_0,z_0)=(-1,-3.48,50)$ as initial condition. After a transient spiking regime, the neuron converges to an irregular, but stable, bursting regime that oscillates around $(\bar{x},\sigma)=(-1,0)$.}
	\label{fig: Alpha5Tau50M85}
\end{figure}

The arrows in Fig.~\ref{fig: Bifurcation} illustrate qualitatively the evolution of the memristive function $\sigma(\bar{x})$, similar to a cobweb plot of a one dimensional map. Consider, for example, an initial value of $\sigma$ greater than zero, as depicted in Fig.~\ref{fig: Bifurcation}(a) prior to the bifurcation. This value of $\sigma$ produces a regime for which the mean value $\bar{x}$, according to Eq.~(\ref{eq: MeanMemristiveSigma}), yields a new value of $\sigma$ smaller than the initial one. In other words, the memristive function decreases because the derivative of Eq.~(\ref{eq: MeanMemristiveSigma}) at $\bar{x}=-1$, given by

\begin{equation}
	\sigma'(-1)=m/2\tau,
	\label{eq: MemristorDerivative}
\end{equation}
\vspace{0.1cm}

\noindent is smaller than $1$ when $m<2\tau$ and the black line is above the blue curve. The process is repeated, making the memristive function $\sigma(\bar{x})$ not to converge to zero, but instead to oscillate around it. The reason for this is that $\sigma=0$ corresponds to a bursting regime when $\alpha=5$ in the original Rulkov model (see Fig.~\ref{fig: ParameterPlane}). During bursts $\bar{x}$ is larger than $-1$ and $\sigma(\bar{x})$ is above zero, whereas during the silent phases between bursts $\bar{x}$ is smaller than $-1$ and $\sigma(\bar{x})$ is below zero, thus producing oscillations around $\sigma=0$. The latter is depicted in Fig.~\ref{fig: Alpha5Tau50M85}(a) when $\tau=50$ and $m=85$. Initially, the memristive function is close to $0.5$, which corresponds to a spiking regime, as Fig.~\ref{fig: Alpha5Tau50M85}(b) shows, but it eventually settles into an irregular bursting regime that oscillates around $(\bar{x},\sigma)=(-1,0)$. Independently from the initial condition, when $m<2\tau$, the neuron converges to a bursting regime around $(\bar{x},\sigma)=(-1,0)$, which can be either periodic or irregular. Therefore, by suitably setting the initial value of $\sigma(\bar{x})$ through $z_0$, uniform transitions from spiking to bursting, as well as from silence to bursting can be obtained.

\subsection{Dynamics of the $\sigma$-memristive Rulkov model after the bifurcation}
\begin{figure}[h!]
	\centering
	\includegraphics[width=\linewidth]{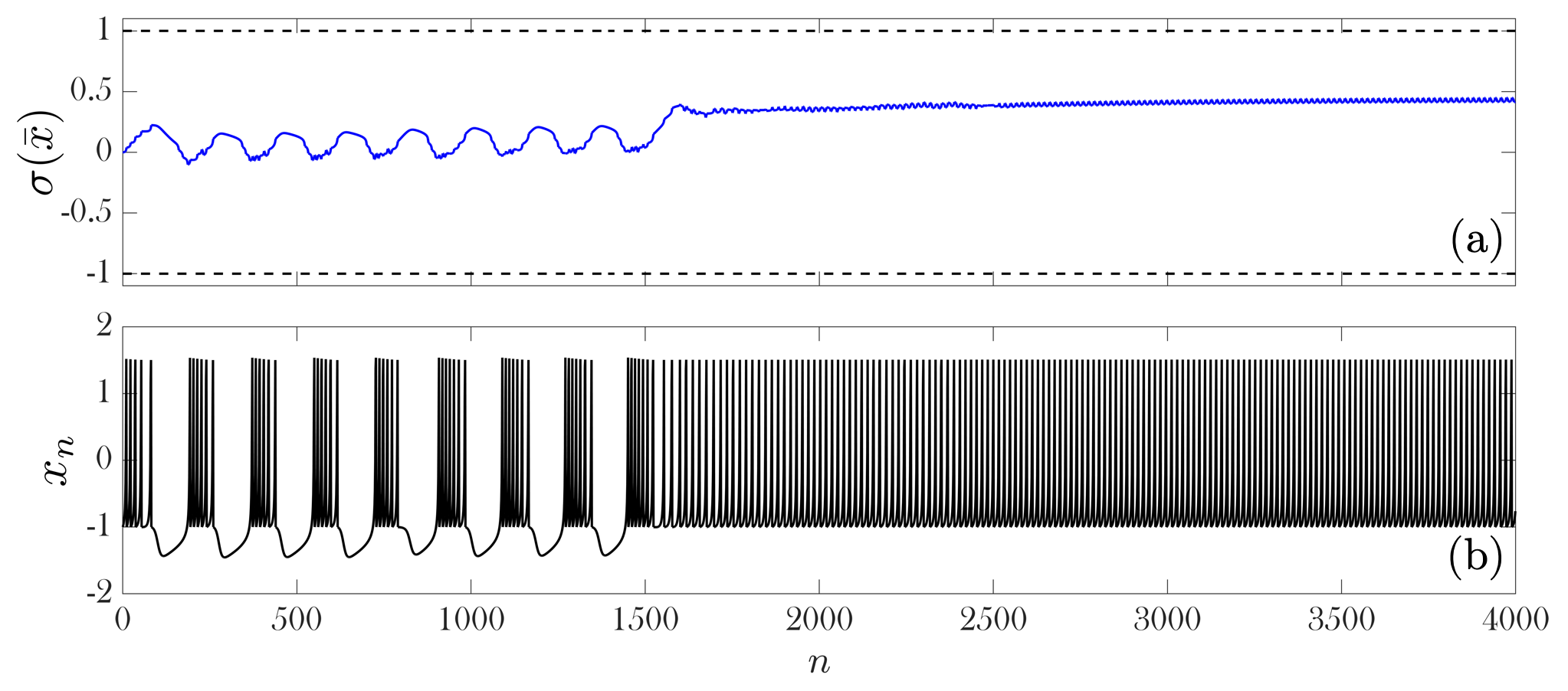}
	\caption{Uniform transition from bursting to spiking after the bifurcation. In (a) we plot the evolution of the memristive function $\sigma(\bar{x})$ and in (b) we plot the evolution of the neuron's voltage $x$ for the $\sigma$-memristive Rulkov model with $\alpha=5$,  $\mu=0.001$, $\tau=70$, $m=150$ and $(x_0,y_0,z_0)=(-1,-3.48,0)$ as initial condition. The system converges to a spiking regime located at $\sigma=0.43$.}
	\label{fig: Alpha5Tau70M150}
\end{figure}

\begin{figure}[h!]
	\centering
	\includegraphics[width=\linewidth]{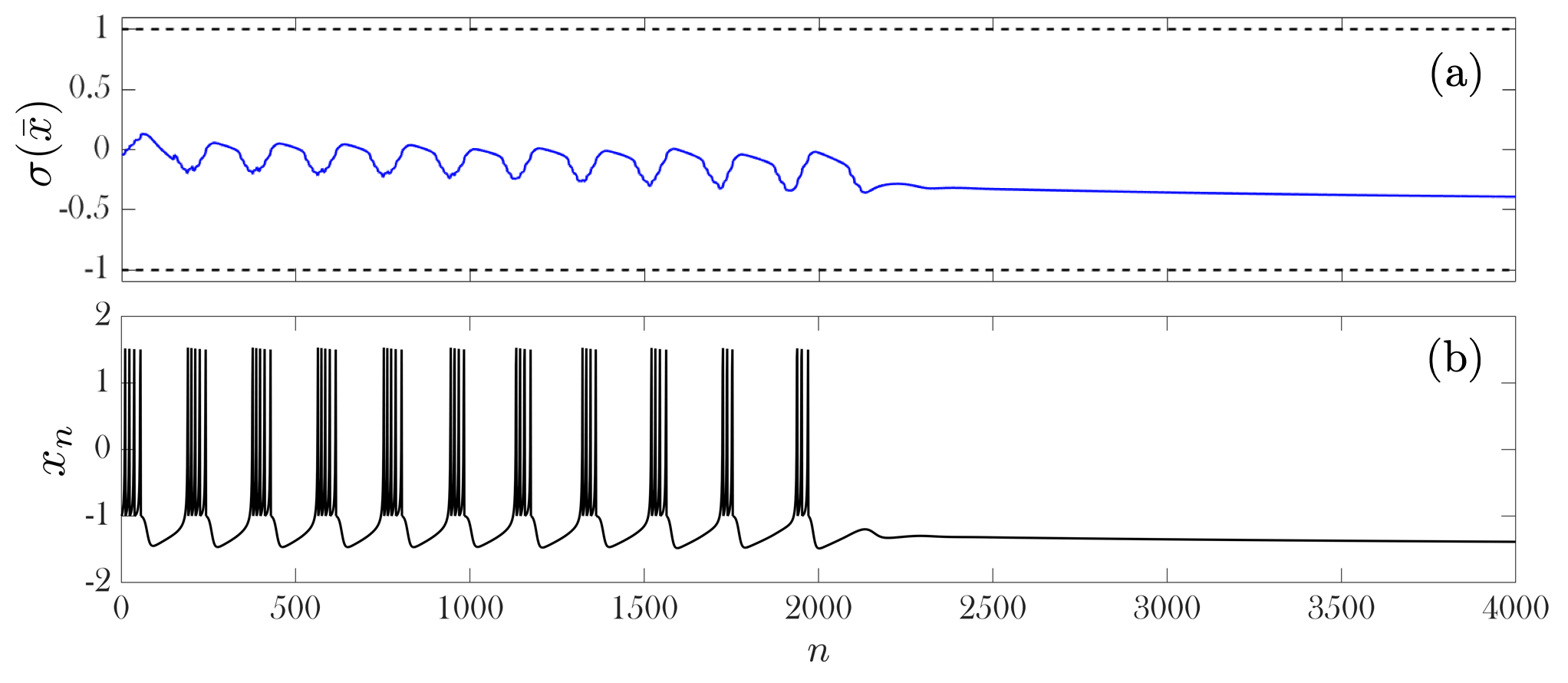}
	\caption{Uniform transition from bursting to silence after the bifurcation. In (a) we plot the evolution of the memristive function $\sigma(\bar{x})$ and in (b) we plot the evolution of the neuron's voltage $x$ for the $\sigma$-memristive Rulkov model with $\alpha=5$, $\mu=0.001$, $\tau=70$, $m=150$ and $(x_0,y_0,z_0)=(-1,-3.48,-6)$ as initial condition. These are the exact same values as in Fig.~\ref{fig: Alpha5Tau70M150}, except for $z_0=-6$, that sets the initial value of the memristive function at $\sigma=-0.043$. The system converges to a silence regime located at $\sigma=-0.43$.}
	\label{fig: Alpha5Tau70M150Z06}
\end{figure}

After the bifurcation, the condition $m>2\tau$ is satisfied, and the stationary bursting regime centered at $(\bar{x},\sigma)=(-1,0)$ becomes unstable because the derivative of the memristive function $\sigma(\bar{x})$ evaluated at this point, given by Eq.~(\ref{eq: MemristorDerivative}), is now greater than $1$. Two new stationary regimes appear after the bifurcation, corresponding to the green dots shown in Fig.~\ref{fig: Bifurcation}(c). The intersection at $\sigma>0$ represents a stable spiking regime, while the intersection located at $\sigma<0$ represents a stable silent regime. The system converges to one of these stable neuronal regimes depending on the initial condition. For instance, both Figs.~\ref{fig: Alpha5Tau70M150} and \ref{fig: Alpha5Tau70M150Z06} show the evolution of the neuron's voltage $x$ and the memristive function when $m=150$ and $\tau=70$. Different final regimes are achieved due to the distinct initial conditions. In Fig.~\ref{fig: Alpha5Tau70M150} the neuron settles into a spiking regime located at $\sigma=0.43$, while in Fig.~\ref{fig: Alpha5Tau70M150Z06} it converges to silent regime located at $\sigma=-0.43$. The stable spiking and silent regimes always occur for $\sigma$ values of opposite sign, due to the odd symmetry of the difference between Eqs.~(\ref{eq: MeanMemristiveSigma}) and (\ref{eq: MeanValueVsSigma}) around $(\bar{x},\sigma)=(-1,0)$.

\subsection{Dynamics of the $\sigma$-memristive Rulkov model at the bifurcation}
\begin{figure}[p]
  	\centering
    	{\includegraphics[width=0.98\linewidth]{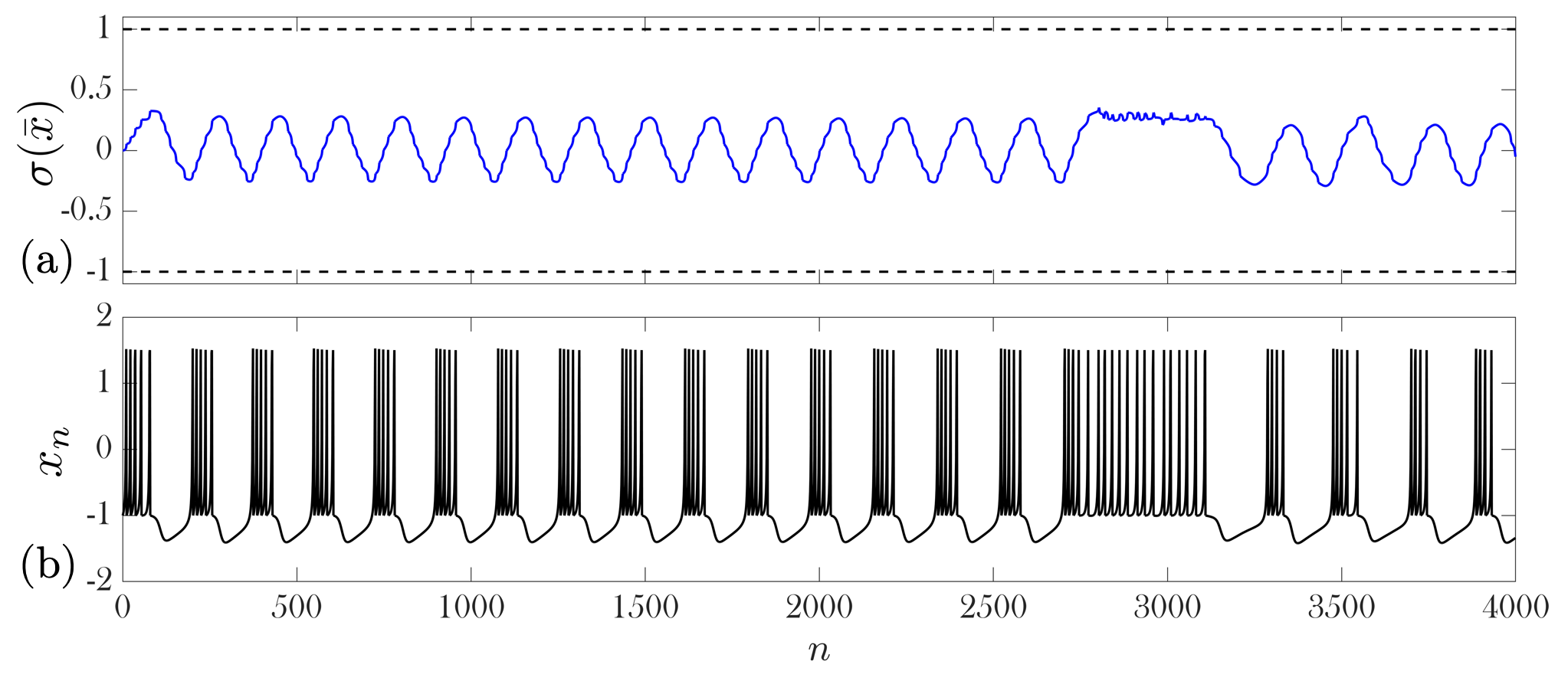}}\vspace{0.1cm}
    	{\includegraphics[width=0.98\linewidth]{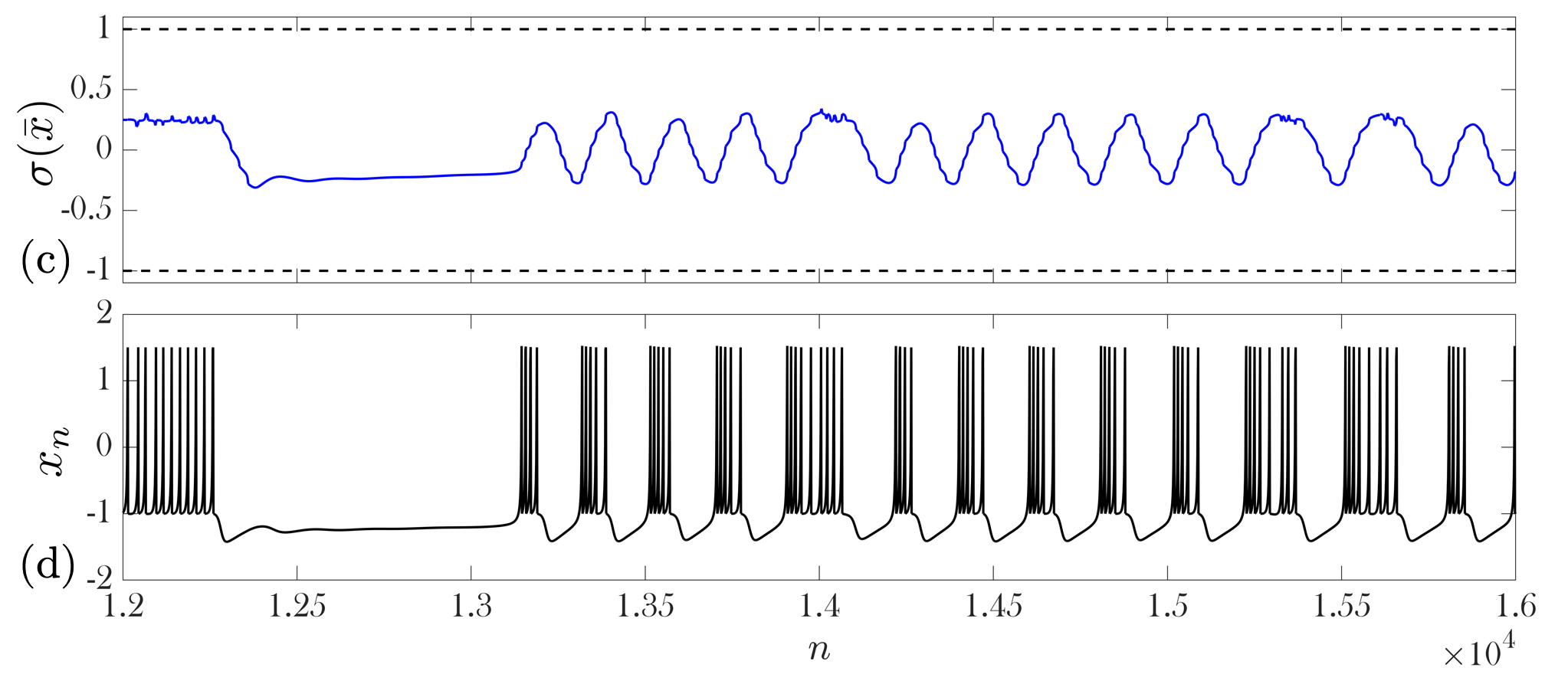}}\vspace{0.1cm}
    	{\includegraphics[width=0.98\linewidth]{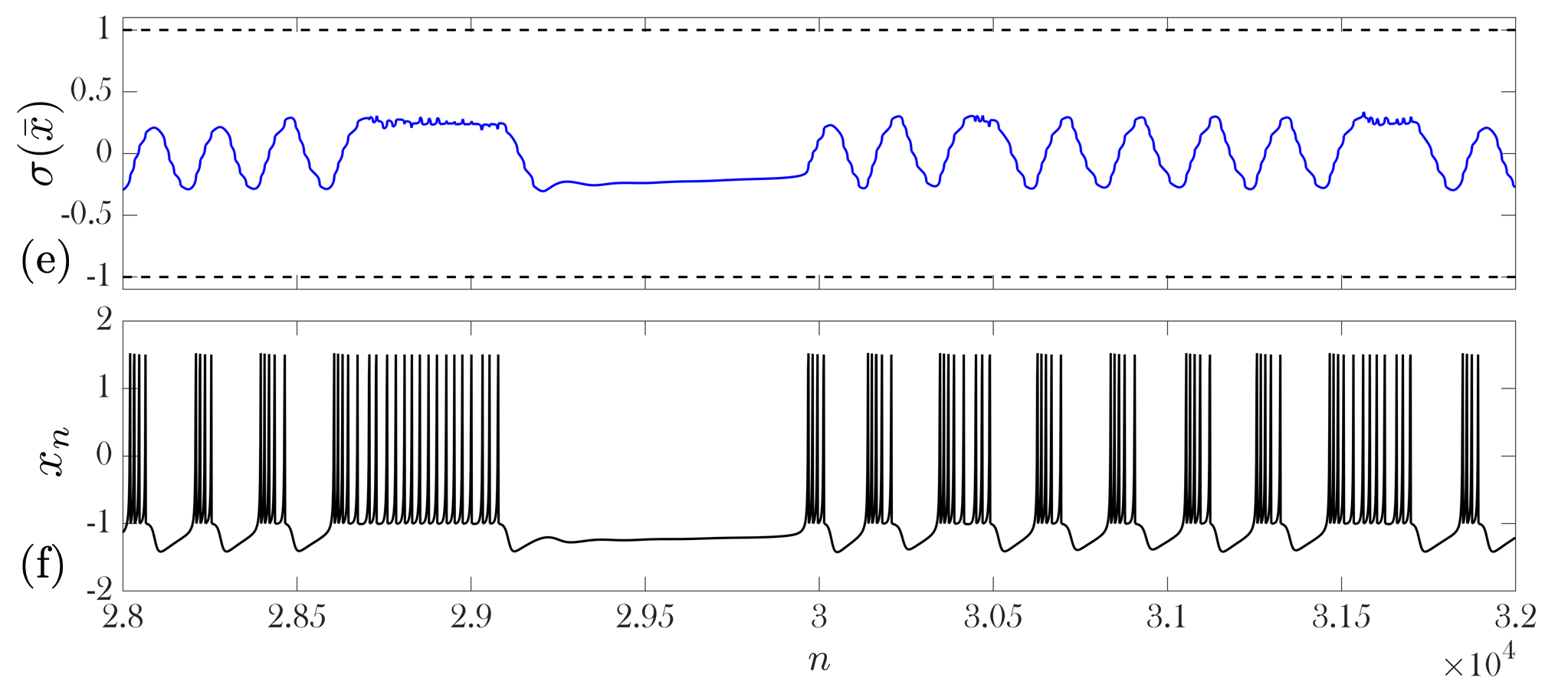}}
	\caption{Evolution of $\sigma(\bar{x})$ (a,c,e) and $x$ (b,d,f) for the $\sigma$-memristive Rulkov model with $\alpha=5$, $\mu=0.001$, $\tau=50$, $m=100$ and $(x_0,y_0,z_0)=(-1,-3.48,0)$ as initial condition. The neuron chaotically switches among the three regimes: silent, spiking and bursting, and never settles down.}
	\label{fig: Alpha5Tau50M100}
\end{figure}

The most interesting dynamics of the newly proposed model arise at the bifurcation, that is, when $m\approx2\tau$ and therefore the derivative of the memristive function at $\bar{x}=-1$ satisfies $\sigma'(-1)\approx1$. Figure~\ref{fig: Alpha5Tau50M100} shows a trajectory of the $\sigma$-memristive Rulkov model with $m=100$ and $\tau=50$, in which the neuron chaotically switches between regimes and never settles into a stationary state. Transitions between neuronal regimes are better characterized than in the original model (compare Fig.~\ref{fig: OriginalTransitions} with Fig.~\ref{fig: Alpha5Tau50M100}), since the discrimination between the ending of a regime and the beginning of a different one become more sharply defined. Additionally, the neuron remains within a given regime for longer intervals. For instance, Fig.~\ref{fig: Alpha5Tau50M100}(e,f) shows that the neuron spends about $1000$ iterations in a silent regime.\\

\begin{figure}[ht!]
  	\centering
    	{\includegraphics[width=\linewidth]{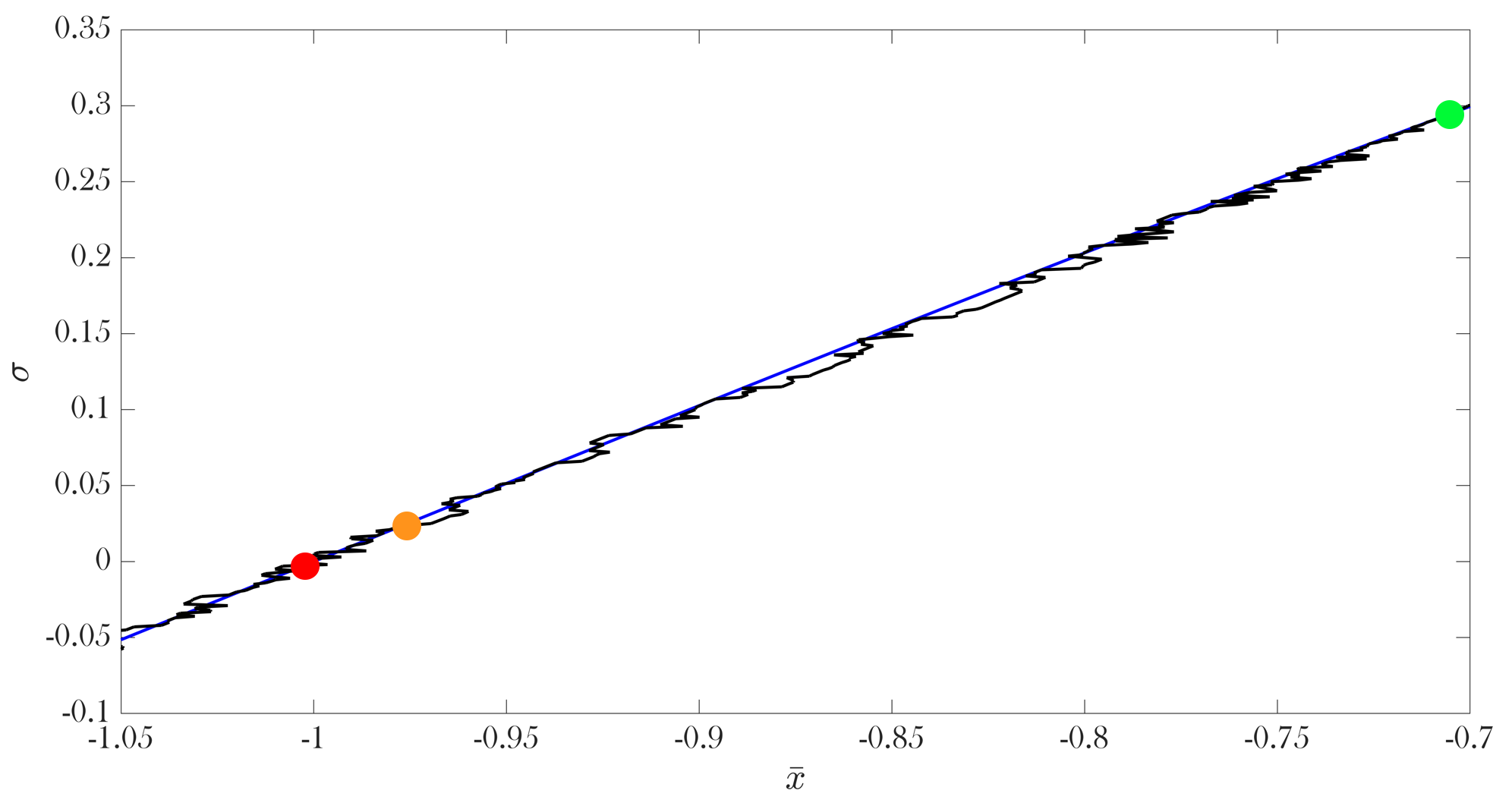}}
	\caption{Bifurcation scenario for $m=103$ and $\tau=50$, enlarged around $(\bar{x},\sigma)=(-1,0)$. The blue cruve represents the memristive function $\sigma(\bar{x})$, while the black curve represents the relation between $\bar{x}$ and $\sigma$ in the original Rulkov model \cite{Rulkov2002}, which deviates considerably from the approximation line $\sigma_0$, given by Eq.~(\ref{eq: MeanValueVsSigma}). These deviations lead the $\sigma$-memristive model to exhibit unexpected stationary regimes, such as the one located at $(-0.9694, 0.0237)$ and marked by an orange dot.}
	\label{fig: LineApproximation}
\end{figure}

\begin{figure}[h!]
  	\centering
    	{\includegraphics[width=\linewidth]{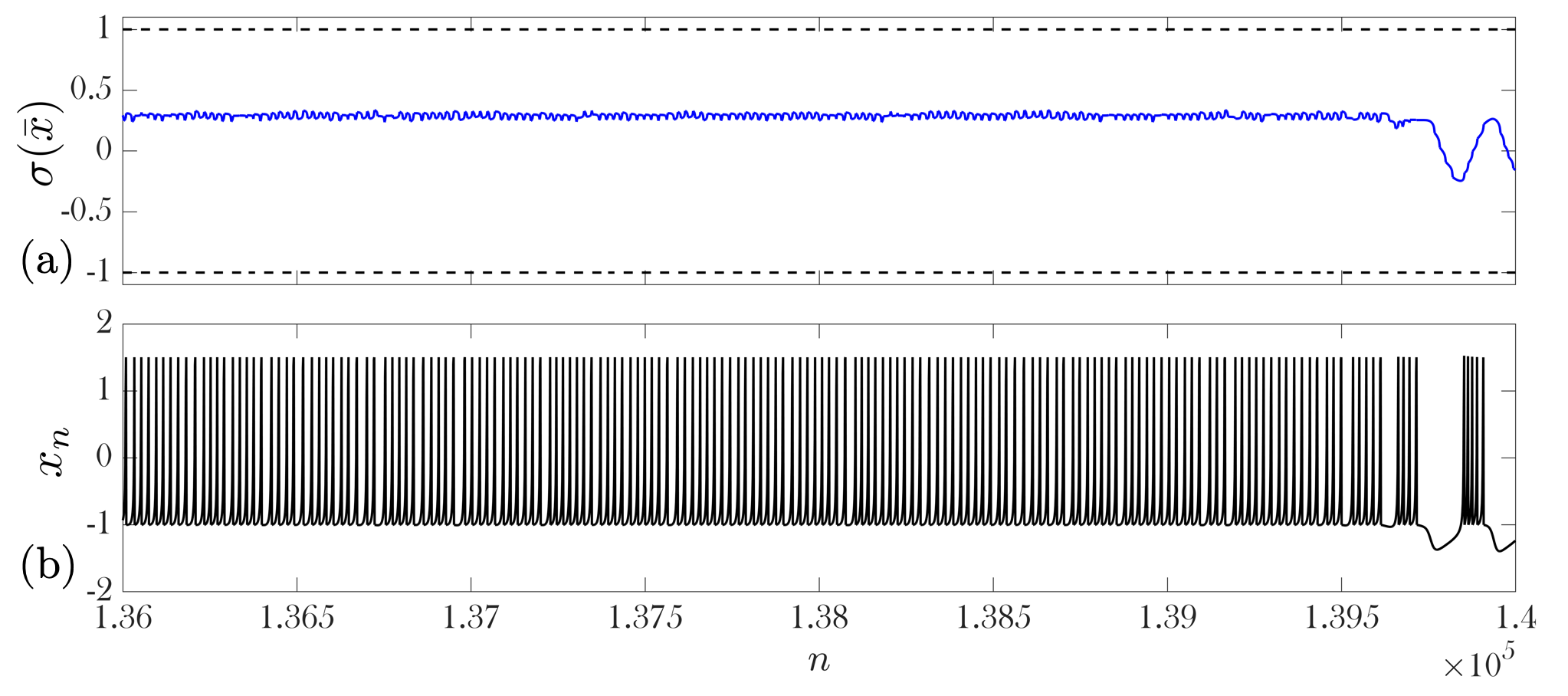}}\vspace{0.1cm}
    	{\includegraphics[width=\linewidth]{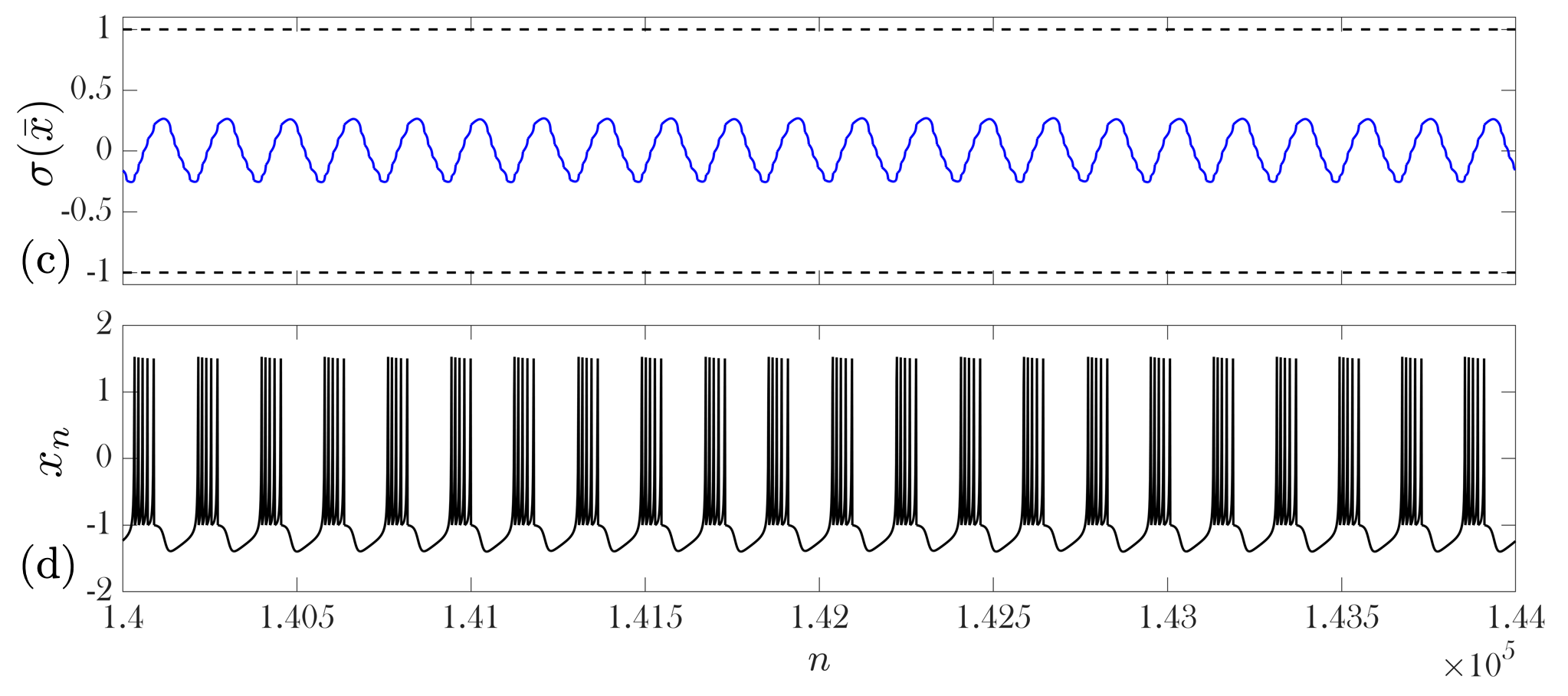}}
	\caption{Same simulation as the one depicted in Fig.~\ref{fig: Alpha5Tau50M100} but with $m=103$ states stored by the neuron. Panels (a,c) show the evolution of the memristive function, and panels (b,d) show the corresponding evolution of the neuron's voltage at the bifurcation scenario for $\alpha=5$, $\mu=0.001$, $\tau=50$, $m=103$ and $(x_0,y_0,z_0)=(-1,-3.48,0)$ as initial condition. The neuron spends about $140000$ iterations at a spiking regime, before ultimately converging to an unexpected regular bursting regime.}
	\label{fig: Alpha5Tau50M103}
\end{figure}

Note that the bifurcation takes place at approximately $m=2\tau$ because Eq.~(\ref{eq: MeanValueVsSigma}) is not completely fulfilled around $(\bar{x},\sigma)=(-1,0)$, as Fig.~\ref{fig: LineApproximation}, which is a zoomed version of Fig.~\ref{fig: Bifurcation}, reveals. As a consequence, just after the bifurcation (i.e., $m\gtrsim2\tau$), additional unexpected stationary regimes might emerge apart from the three illustrated in Fig.~\ref{fig: Bifurcation}(c), such as the one indicated by an orange dot in Fig.~\ref{fig: LineApproximation}. For instance, Fig.~\ref{fig: Alpha5Tau50M103} shows an excerpt of the $\sigma$-memristive model evolution with $m=103$ and $\tau=50$, that has the stable regimes located at $\sigma=\pm0.2921$, with the positive value corresponding to a spiking regime indicated by the green dot in Fig.~\ref{fig: LineApproximation}. The memristive neuron spends about $140000$ iterations spiking and never reaches the steady regime located at $\sigma=0.2921$. It eventually converges to an unexpected regular bursting regime, that corresponds to the orange dot in Fig.~\ref{fig: LineApproximation}.\\

The $\sigma$-memristive Rulkov model has been shown for $\alpha=5$. For other values of $\alpha$, the model still presents the three distinct scenarios. Nevertheless, the stable spiking and silent regimes originated after the bifurcation need greater values of $m/\tau$ to form as $\alpha$ increases. This is because, when increasing $\alpha$, the bursting region is enlarged, as portrayed in Fig.~\ref{fig: ParameterPlane}, and the new stationary regimes formed are also of bursting. The most interesting dynamics emerges for values of $\alpha$ close to $5$, for which transitions between neuronal regimes are found at the bifurcation $m\approx2\tau$. We also note that our analysis of the model based on the mean value of $x$ requires large values of $m$.\\

Other memristive functions $\sigma(\bar{x})$, in addition to Eq.~(\ref{eq: MeanMemristiveSigma}), could be chosen to intersect the line $\sigma_0$ at different points, thus creating other stationary regimes. For instance, following the scheme used in Ref.~\cite{NovelChaoticMap} for the process equation, one could construct a memristive function that intersects the line $\sigma_0$ multiple times, with each intersection corresponding to a stationary regime of unexpected stability.

\section{Conclusions}
\label{sec: Conclusions}
In this work, we have modified the Rulkov model by replacing one of its two control parameters by a memristive function in order to generate transitions between different neuronal regimes. We have shown that the parameter $\alpha$ must not be used as a memristive function because, in doing so, it can interfere negatively with some of the essential properties of the original model. On the other hand, turning $\sigma$ into a memristive function with finite memory produces interesting, both uniform and chaotic, transitions among neuronal regimes while it preserves the biological features of the model.\\

The newly proposed $\sigma$-memristive Rulkov model exhibits three possible scenarios. Prior to the bifurcation, uniform transitions to a stable bursting regime are achieved. At the bifurcation, complex transitions between distinct neuronal regimes, that last long periods of time, tend to emerge. After the bifurcation, uniform transitions to a stable spiking or silent regime are found, depending on the initial state of the neuron. The scenario of the model is selected by properly setting the relation between the number of states $m$ stored by the neuron and the rate of change $\tau$ of the memristive function. Therefore, we can conclude that the dynamics are influenced by the neuron's memory.\\

Both the original Rulkov model and its $\sigma$-memristive version reproduce a particular type of bursting ~\cite{BurstingFeatures}. A natural next step would be to apply the strategy developed in this work to other map-based models exhibiting different bursting features. Moreover, the proposed $\sigma$-memristive model could be used to fit physiological experimental data, in parallel with the original Rulkov model, by employing some of the techniques described in Ref.~\cite{NeuronOptimization}, possibly incorporating rare-event analysis as in Ref.~\cite{RareEventPrediction}. Finally, it would be interesting to consider a time-varying neuronal memory, namely a non-constant number of stored states $m$. Varying $m$ would allow a dynamic combination of the three distinct scenarios presented in this paper.

\section*{CRediT authorship contribution statement}

Miguel Moreno: Conceptualization, Formal analysis, Investigation, Methodology, Software, Validation, Visualization, Writing – original draft, Writing – review \& editing. Alexandre R. Nieto: Conceptualization, Formal analysis, Investigation, Methodology, Software, Supervision, Validation, Writing – review \& editing. Miguel A.F. Sanjuán: Conceptualization, Formal analysis, Funding acquisition, Investigation, Methodology, Project administration, Supervision, Validation, Writing – review \& editing.

\section*{Declaration of competing interest}
All authors declare that they have no known competing financial interests or personal relationships that could have appeared to influence the work reported in this paper.

\section*{Acknowledgments}

Financial support by the Spanish State Research Agency (AEI) and the European Regional Development Fund (ERDF, EU) under Project No.~PID2023-148160NB-I00 (MCIN/AEI/10.13039/501100011033) is acknowledged.

\section*{Data availability}
No data was used for the research described in the article.

\bibliographystyle{elsarticle-num}
\bibliography{bibliografia}

\end{document}